\documentclass{pasj00}

\begin{document}
\SetRunningHead{T.~Morokuma et al.}{KISS Survey Strategy}
\Received{2014/07/10}
\Accepted{2014/08/26}

\title{
Kiso Supernova Survey (KISS): Survey Strategy
}

%

\author{
   Tomoki \textsc{Morokuma}\altaffilmark{1},     
   Nozomu \textsc{Tominaga}\altaffilmark{2,3},     
   Masaomi \textsc{Tanaka}\altaffilmark{4},      
   Kensho \textsc{Mori}\altaffilmark{5}, 
   Emiko \textsc{Matsumoto}\altaffilmark{2}, \\
   Yuki \textsc{Kikuchi}\altaffilmark{1}, 
   Takumi \textsc{Shibata}\altaffilmark{2}, 
   Shigeyuki \textsc{Sako}\altaffilmark{1}, 
   Tsutomu \textsc{Aoki}\altaffilmark{6}, 
   Mamoru \textsc{Doi}\altaffilmark{1,7},      
   Naoto \textsc{Kobayashi}\altaffilmark{1}, 
   Hiroyuki \textsc{Maehara}\altaffilmark{6}, 
   Noriyuki \textsc{Matsunaga}\altaffilmark{8}, 
   Hiroyuki \textsc{Mito}\altaffilmark{6}, 
   Takashi \textsc{Miyata}\altaffilmark{1}, 
   Yoshikazu \textsc{Nakada}\altaffilmark{1}, \\
   Takao \textsc{Soyano}\altaffilmark{6}, 
   Ken'ichi \textsc{Tarusawa}\altaffilmark{6}, 
   Satoshi \textsc{Miyazaki}\altaffilmark{4}, 
   Fumiaki \textsc{Nakata}\altaffilmark{9}, 
   Norio \textsc{Okada}\altaffilmark{4}, \\
   Yuki \textsc{Sarugaku}\altaffilmark{10}, 
   Michael W. \textsc{Richmond}\altaffilmark{11}, 
   Hiroshi \textsc{Akitaya}\altaffilmark{12},      
   Greg \textsc{Aldering}\altaffilmark{13}, \\
   Ko \textsc{Arimatsu}\altaffilmark{10,8},      
   Carlos \textsc{Contreras}\altaffilmark{14,15},      
   Takashi \textsc{Horiuchi}\altaffilmark{16},      
   Eric Y. \textsc{Hsiao}\altaffilmark{14,15},      
   Ryosuke \textsc{Itoh}\altaffilmark{5},      \\
   Ikuru \textsc{Iwata}\altaffilmark{9},      
   Koji, S. \textsc{Kawabata}\altaffilmark{12},      
   Nobuyuki \textsc{Kawai}\altaffilmark{17},      
   Yutaro \textsc{Kitagawa}\altaffilmark{1},      
   Mitsuru \textsc{Kokubo}\altaffilmark{1},      \\
   Daisuke \textsc{Kuroda}\altaffilmark{18},     
   Paolo, \textsc{Mazzali}\altaffilmark{19,20,21},  
   Toru \textsc{Misawa}\altaffilmark{22},      
   Yuki \textsc{Moritani}\altaffilmark{12},      
   Nidia \textsc{Morrell}\altaffilmark{14},     \\ 
   Rina \textsc{Okamoto}\altaffilmark{16},  
   Nikolay \textsc{Pavlyuk}\altaffilmark{23}, 
   Mark M. \textsc{Phillips}\altaffilmark{14},  
   Elena \textsc{Pian}\altaffilmark{24,25}, 
   Devendra \textsc{Sahu}\altaffilmark{26}, \\
   Yoshihiko \textsc{Saito}\altaffilmark{17},
   Kei \textsc{Sano}\altaffilmark{10,8},   
   Maximilian D. \textsc{Stritzinger}\altaffilmark{15},      
   Yutaro \textsc{Tachibana}\altaffilmark{17},      \\
   Francesco \textsc{Taddia}\altaffilmark{27},      
   Katsutoshi \textsc{Takaki}\altaffilmark{5},      
   Ken \textsc{Tateuchi}\altaffilmark{1},           
   Akihiko \textsc{Tomita}\altaffilmark{28},  \\  
   Dmitry \textsc{Tsvetkov}\altaffilmark{23}, 
   Takahiro \textsc{Ui}\altaffilmark{5},      
   Nobuharu \textsc{Ukita}\altaffilmark{18},      
   Yuji \textsc{Urata}\altaffilmark{29}, \\
   Emma S. \textsc{Walker}\altaffilmark{30},      
   Taketoshi \textsc{Yoshii}\altaffilmark{17}
   }

\altaffiltext{1}{Institute of Astronomy, Graduate School of Science, The University of Tokyo, 2-21-1, Osawa, Mitaka, Tokyo 181-0015, Japan}
\altaffiltext{2}{Department of Physics, Faculty of Science and Engineering, Konan University, 8-9-1 Okamoto, Kobe, Hyogo 658-8501, Japan}
\altaffiltext{3}{Kavli Institute for the Physics and Mathematics of the Universe (WPI), The University of Tokyo, 5-1-5 Kashiwanoha, Kashiwa, Chiba 277-8583, Japan}
\altaffiltext{4}{National Astronomical Observatory of Japan, 2-21-1, Osawa, Mitaka, Tokyo 181-8588, Japan}
\altaffiltext{5}{Department of Physical Science, Hiroshima University, Higashi-Hiroshima, Hiroshima 739-8526, Japan}
\altaffiltext{6}{Kiso Observatory, Institute of Astronomy, Graduate School of Science, The University of Tokyo 10762-30, Mitake, Kiso-machi, Kiso-gun, Nagano 397-0101, Japan}
\altaffiltext{7}{Research Center for the Early Universe, Graduate School of Science, The University of Tokyo, 7-3-1 Hongo, Bunkyo-ku, Tokyo 113-0033}
\altaffiltext{8}{Department of Astronomy, Graduate School of Science, The University of Tokyo, Hongo 7-3-1, Bunkyo-ku, Tokyo 113-0033, Japan}
\altaffiltext{9}{Subaru Telescope, 650 North A’ohoku Place, Hilo, HI 96720, USA}
\altaffiltext{10}{Institute of Space and Astronautical Science, Japan Aerospace Exploration Agency, 3-1-1 Yoshinodai, Chuo-ku, Sagamihara, Kanagawa 252-5210, Japan}
\altaffiltext{11}{Department of Physics, Rochester Institute of Technology, Building 76-1274, 85 Lomb Memorial Drive, Rochester, NY 14623-5603, USA}
\altaffiltext{12}{Hiroshima Astrophysical Science Center, Hiroshima University, Higashi-Hiroshima, Hiroshima 739-8526, Japan}
\altaffiltext{13}{Lawrence Berkeley National Laboratory, 1 Cyclotron Road, Berkeley, CA 94720, USA}
\altaffiltext{14}{Carnegie Observatories, Las Campanas Observatory, Colina El Pino, Casilla 601, Chile}
\altaffiltext{15}{Department of Physics and Astronomy, Aarhus University, Ny Munkegade, DK-8000 Aarhus C, Denmark}
\altaffiltext{16}{Department of Physics, Faculty of Science, Shinshu University, 3-1-1 Asahi, Matsumoto, Nagano 390-8621, Japan}
\altaffiltext{17}{Department of Physics, Tokyo Institute of Technology, 2-12-1 Ookayama, Meguro-ku, Tokyo 152-8551, Japan}
\altaffiltext{18}{Okayama Astrophysical Observatory, National Astronomical Observatory of Japan, 3037-5 Honjo, Kamogata, Asakuchi, Okayama 719-0232, Japan}
\altaffiltext{19}{Astrophysics Research Institute, Liverpool John Moores University, IC2, Liverpool Science Park, 146 Brownlow Hill, Liverpool L3 5RF, UK}
\altaffiltext{20}{Max-Planck Institut f$\ddot{u}$r Astrophysik, Karl-Schwarzschildstr. 1, D-85748 Garching, Germany}
\altaffiltext{21}{INAF-Osservatorio Astronomico di Padova, Vicolo dell'Osservatorio 5, I-35122 Padova, Italy}
\altaffiltext{22}{School of General Education, Shinshu University, 3-1-1 Asahi, Matsumoto, Nagano 390-8621, Japan}
\altaffiltext{23}{Lomonosov Moscow State University, Sternberg Astronomical Institute, 13 Universitetskij prospekt, Moscow 119234, Russia}
\altaffiltext{24}{Scuola Normale Superiore di Pisa, Piazza dei Cavalieri 7, I-56126 Pisa, Italy}
\altaffiltext{25}{INAF-Istituto di Astrofisica Spaziale e Fisica Cosmica, Via P. Gobetti 101, I-40129 Bologna, Italy}
\altaffiltext{26}{Indian Institute of Astrophysics, Koramangala, Bangalore 560 034, India}
\altaffiltext{27}{The Oskar Klein Centre, Department of Astronomy, Stockholm University, AlbaNova, SE-10691 Stockholm, Sweden}
\altaffiltext{28}{Faculty of Education, Wakayama University, Sakaedani 930, Wakayama 640-8510, Japan}
\altaffiltext{29}{Institute of Astronomy, National Central University, Chung-Li 32054, Taiwan}
\altaffiltext{30}{Department of Physics, Yale University, PO Box 208120, New Haven, CT 06520-8120, USA}
\email{tmorokuma@ioa.s.u-tokyo.ac.jp}

%


\maketitle

\begin{abstract}
The Kiso Supernova Survey (KISS) is a high-cadence optical wide-field supernova (SN) survey. 
The primary goal of the survey is to catch the very early light of a SN, during the shock breakout phase.
Detection of SN shock breakouts combined with multi-band photometry 
obtained with other facilities would provide detailed physical information on the progenitor stars of SNe. 
The survey is performed using a $2.2\times2.2$~deg field-of-view instrument
on the 1.05-m Kiso Schmidt telescope, the Kiso Wide Field Camera (KWFC). 
We take a three-minute exposure in $g$-band once every hour in our survey, reaching magnitude $g \sim 20-21$.
About 100~nights of telescope time per year have been spent on the survey since April 2012. 
The number of the shock breakout detections is estimated to be of order of 1 during our 3-year project. 
This paper summarizes the KISS project including 
the KWFC observing setup, the survey strategy, the data reduction system, and CBET-reported SNe discovered so far by KISS. 
\end{abstract}

\section{Introduction}
\label{sec:intro}
The variable 
sky has been intensively explored 
by wide-field surveys such as 
Palomar Transient Factory (PTF; \cite{rau2009}; \cite{law2009}), 
Catalina Real-Time Sky Survey (CRTS; \cite{drake2009}), 
La~Silla-QUEST Low Redshift Supernova Survey \citep{baltay2013}, 
Mobile Astronomical System of the TElescope-Robots (MASTER; \cite{lipunov2004}), 
Panoramic Survey Telescope \& Rapid Response System (Pan-STARRS1; \cite{kaiser2002}), 
SkyMapper \citep{keller2007}, 
SDSS Stripe~82 Supernova Survey (\cite{frieman2008}; \cite{sako2008}; \cite{sako2014}), 
and 
Deep Lens Survey (DLS; \cite{wittman2002}),
with each survey having a uniform data quality controlled systematically, 
especially during the last two decades. 
Some of the ground-based large aperture and space telescopes also have utilized 
their deep imaging capabilities to explore transient phenomena 
within the scheme of their deep surveys 
(\cite{sarajedini2003}; \cite{sarajedini2006}; \cite{cohen2006}; 
\cite{morokuma2008}; \cite{villforth2010}; \cite{grogin2011}; \cite{postman2012}). 
All these projects were made possible by the recent development of 
large mosaiced-CCD imaging cameras. 
However, 
we still have little knowledge of fast transient objects  
whose time scales are shorter than a day except for a few classes of objects
(e.g., rapidly variable stars such as RR~Lyrae).  
One of the most interesting phenomena in this time scale is the shock breakout of a supernova (SN). 
Although the existence of this phenomenon was theoretically 
predicted about 40~years ago \citep{klein1978} and many SN surveys have been conducted, 
there are only three serendipitous detections so far 
(SN~2008D in NGC~2770 in X-ray; \cite{soderberg2008}; 
SNLS-04D2dc and SNLS-06D1jd in ultraviolet; \cite{schawinski2008,gezari2008}) 
due to the short time scale: just a few hours. 
Shock breakouts are one of the brightest phenomena associated with SNe and 
are considered to be associated with every SN. 
Shock breakouts of SN explosions with hydrogen envelopes in particular are luminous 
in the optical and last for a few hours 
by virtue of the large radius of a progenitor star. 
They are expected to be a new tool for exploring the distant universe \citep{tominaga2011}. 
In this paper, we show our new survey optimized for detecting nearby SN shock breakouts. 

A new wide-field optical imager, the Kiso Wide Field Camera (KWFC; \cite{sako2012}), 
for the 1.05-m Kiso Schmidt telescope 
operated by the Institute of Astronomy of the University of Tokyo, 
in Nagano, Japan, was developed and has been open to public use 
since April 2012, succeeding the former camera 2kCCD \citep{itoh2001}.
We started a high-cadence SN survey, called the Kiso Supernova Survey~(KISS), 
with KWFC in April 2012 within the scheme of the Kiso Observatory Large Programs. 
Another program is a search mainly for periodic variable stars in the Galactic plane, 
called the KWFC Intensive Survey Of the Galactic Plane~(KISOGP). 
These programs are scheduled to be conducted until March 2015. 

The structure of this paper is as follows. 
We summarize the KISS observations 
including our observing strategy and 
the data reduction system 
in \S\ref{sec:observingstrategy} and \S\ref{sec:datareductiontransientdetection}, respectively. 
Our initial results are shown in \S\ref{sec:initialresults}. 
\S\ref{sec:summary} is a summary of the paper.
We use the standard $\Lambda$CDM cosmological parameters of 
$(H_0, \Omega_M, \Omega_\Lambda)=(71, 0.27, 0.73)$ \citep{komatsu2011}. 
All magnitudes are measured in the AB system.
All coordinates are measured in J2000 and all the dates are in UT. 

\section{KISS Observations}
\label{sec:observingstrategy}

\subsection{Kiso Wide Field Camera (KWFC)}
\label{sec:kwfcperformance}
\begin{figure}
  \begin{center}
    \includegraphics[width=80mm]{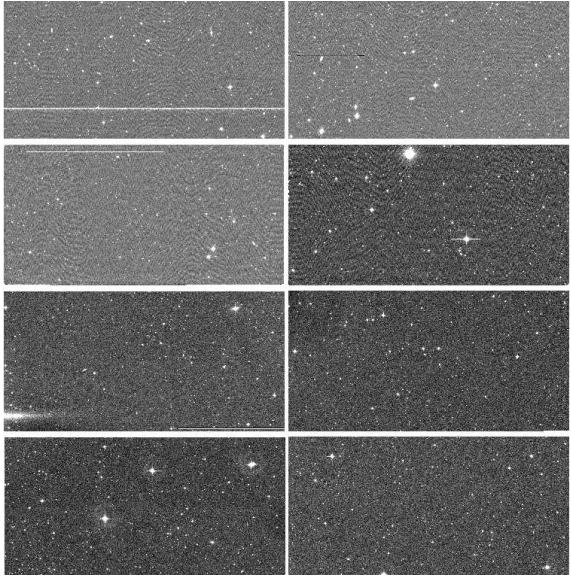}
  \end{center}
  \caption{
An example of the entire KWFC field-of-view 
($1\times1$ binning, $g$-band, 300 sec exposure). 
The top four CCDs are made by SITe and the bottom four CCDs those of MIT/LL. 
North is up and east is left. 
The gaps on sky are 82-84~arcsec and 48-62~arcsec in the north-south and east-west directions, respectively. 
There are some defects on the CCDs.
}
\label{fig:kwfcimage}
\end{figure}

The KWFC is an optical wide-field 64~megapixel imager attached to the prime focus of the Kiso Schmidt telescope 
covering $2.2\times2.2$ deg 
with 4~SITe and 
4~MIT Lincoln Laboratory (MIT/LL) 
15$\mu$m 2k$\times$4k CCDs (Figure~\ref{fig:kwfcimage}). 
The plate scale is 0.946~arcsec pixel$^{-1}$ 
and the effective area is 4.6~deg$^2$. 
KWFC is uniquely equipped with 
a filter exchanger based on an industrial robot arm and a magazine unit capable of storing 12~filters.
For the KISS observations, we adopt $1\times1$ binning (no binning) 
and SLOW (all 8-chip) readout mode. 
All the CCDs are read out by the Kiso Array Controller (KAC; \cite{sako2012}). 
In this mode, the readout noise is about 20 and $5-10$ electrons for the SITe and MIT/LL CCDs, respectively. 
The quantum efficiency of these CCDs is 40-65\% in the $g$-band used for the KISS observations. 
The readout time in total is 125 sec including the wiping time. 
Typical telescope slewing time from field to field, which is usually finished during the readout, 
is about 2~minutes. 
In total, one typical 3-minute exposure takes 
about 5-6~minutes and the observing efficiency is about 50-60\%. 
These parameters, as well as the survey strategy of KISS, are summarized in Table \ref{tab:kisssurveystrategy}.

\begin{table}
  \caption{
Summary of the KISS survey instrument and strategy. 
  }\label{tab:kisssurveystrategy}
  \begin{center}
\scriptsize{
    \begin{tabular}{ll}
      \hline
      telescope & 1.05-m Kiso Schmidt\\
      instrument & Kiso Wide Field Camera (KWFC)\\
      detector & 4 MIT/LL CCDs and 4 SITe CCDs (15$\mu$m pixel)\\
      field-of-view & 2.2~deg~$\times$~2.2~deg\\
      pixel scale & 0.946 arcsec pixel$^{-1}$\\
      effective area & 4.6~deg$^{2}$\\
      pixel binning & $1\times1$ (none)\\
      filter & $g$-band\\
      exposure time & 180 sec\\
      cadence & $\sim1$~hour\\
      survey field & SDSS fields with high star formation rates\\
      time allocation & $\sim100$~nights per year, around new moon\\
      \hline
    \end{tabular}
}
  \end{center}
\end{table}

\subsection{Kiso/KWFC Observing Setup, Filter Selection, and Cadence}
\label{sec:observingsetup}

The primary purpose of KISS is to catch the very early light of mainly 
core-collapse SN, the shock breakout phase, whose rising and declining time scales are as short as a few hours. 
The fact that there have been only 3 serendipitous detections of 
shock breakouts at other wavelengths 
\citep{soderberg2008,schawinski2008,gezari2008} 
indicates that the
detection of shock breakouts requires a specially optimized 
high-cadence wide-field systematic survey. 
A successful survey must also 
have quick data reduction 
and quick follow-up observations for identification and characterization. 

First, we conduct a single-band survey in $g$-band. 
Observations at short wavelengths are more effective in catching shock breakout light 
because the spectrum of the shock breakout can be approximated by quasi-blackbody radiation and 
the color temperature is of order 100,000~K. 
Therefore, the SED peaks in the far-UV and 
the optical wavelength region is in the Rayleigh-Jeans regime:
the flux density 
rapidly decreases with wavelength~($f_{\lambda}\propto\lambda^{-4}$). 
We choose $g$-band due to the higher total throughput of this band 
compared to the shorter-wavelength $u$-band. 
Second, we adopt a high cadence, as short as 1~hour, based on our detailed light curve simulations
done in \S\ref{sec:numberestimate}. 
As seen in \S\ref{sec:numberestimate} and previous papers (\cite{schawinski2008,gezari2008}), 
the time scales of the luminosity changes are as short as a few hours 
and observational information in this time scale would be important for deriving the physical parameters 
of shock breakout phenomenon. 

KISS survey regions are selected from SDSS imaging fields 
where spectroscopic diagnostics for star formation rates 
are available\footnote{The MPA-JHU DR7 release of spectrum measurements. 
http://www.mpa-garching.mpg.de/SDSS/DR7/}\citep{brinchmann2004},
which are used to estimate the total star formation rates per KWFC field-of-view (FoV). 
The typical total star formation rate per KWFC FoV in our survey field is 
$\sim20$ M$_\odot$ yr$^{-1}$ at $z<0.03$. 
Archival images from the SDSS taken several years ago are used 
as reference images, since they are deeper and have higher spatial resolution than the KWFC images. 
Considering the seeing statistics in $g$-band at the Kiso site 
(typically 3.3-5.3~arcsec FWHM at 20-80~percentile; 3.9~arcsec FWHM at median),
we adopt $1\times1$~binning 
in order to construct finely sampled point spread functions in the KWFC images 
for image subtraction with which transient and variable objects are isolated and located. 
We also choose the SLOW (all the 8 KWFC chips) 
readout mode to cover as wide an area as possible\footnote{Other options for the binning and readout are $2\times2$ binning and FAST 
where only 4 MIT/LL CCDs are read out, respectively.}. 

\begin{figure}
  \begin{center}
    \includegraphics[angle=270,width=80mm]{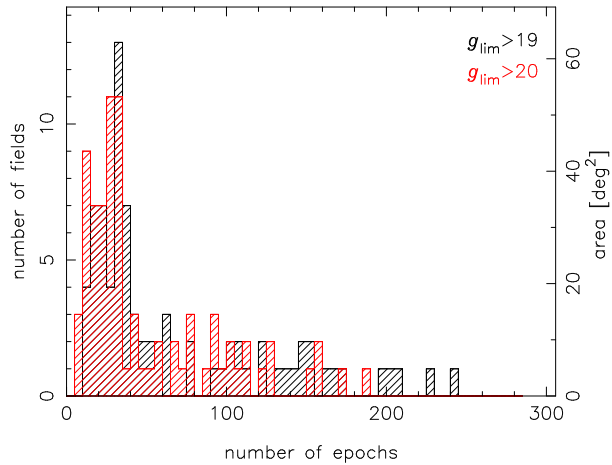}
  \end{center}
  \caption{
  The number of survey fields as a function of the number of the KISS observation epochs for different limiting magnitudes, 
  $g_{\rm{lim}}>19$~mag in black and $g_{\rm{lim}}>20$~mag in red. 
}
\label{fig:numberofepochs}
\end{figure}

In a typical case, we repeat a cycle which consists of 
one 3-minute exposure for $20$ different regions 
about 5 times per night. 
The depth of each image, defined as 50\% detection completeness estimated by embedding artificial objects 
using IRAF/{\it artdata}, is $g\sim20-21$ mag. 
The KISS detection limit matches the peak brightness of a typical shock breakout 
for a star of initial mass $20~M_{\rm{\odot}}$, explosion energy $E=1.0\times10^{51}$~[erg] 
at a distance of 160~Mpc (corresponding to $z\sim0.04$), 
at which the SDSS spectroscopy completeness limit ($r<17.7$) covers 
most of the populations of star-forming galaxies with supernova detection 
in the PTF~\citep{arcavi2010}\footnote{The assumed maximum distance of $d=160$~Mpc corresponds 
to a distance modulus $DM=36.0$~mag; therefore, $r<17.7$~mag corresponds to $M_r<-18.3$~mag. 
In \citet{arcavi2010}, 
most of the core-collapse SNe occur in galaxies brighter than this magnitude limit.}. 
In total, we observe about 100~deg$^2$ per night. 
The distribution of the number of epochs is shown in Figure~\ref{fig:numberofepochs}. 
A considerable number of fields have been visited in more than 100 epochs due to the high frequency of several visits per night. 
Under cloudy weather conditions, which provide shallower images 
up to $g<19$, we change the strategy and observe several 
pointings for nearby galaxy clusters and groups. 

About 100 nights per year (about 10~nights per month around new moon) are used for KISS observations;
roughly half of them shared with other science programs. 
Every year, from mid-June to mid-August, when weather at the site is
affected by the Japanese rainy season, the telescope and camera are under maintenance.

\subsection{Follow-Up Observation Strategy}
\label{sec:followupobs}
Optical multi-band light curves are required to establish our physical understanding 
of a shock breakout \citep{tominaga2011} and thus 
it is important to launch follow-up imaging observations right after the discovery. 
When we find a candidate for shock breakout 
after the procedure described in \S\ref{sec:datareductiontransientdetection} including 
quick automatic data reduction, image subtraction, candidate detection, and 
target screening, 
we trigger the Akeno~50-cm/MITSuME \citep{kotani2005,yatsu2007,shimokawabe2008} 
for simultaneous $g'$-, $R_C$-, and $I_C$-band 
automatic imaging observations for confirmation. 
The development of this automatic follow-up system is 
based on an automatic observation system originally developed for GRB optical afterglow follow-ups 
with MITSuME, which was completed in June 2014. 
The MITSuME images, for which we do not perform image subtraction, are helpful 
if the SNe are spatially separated from the host galaxies or are on the diffuse host galaxies. 
When the candidate's existence is confirmed in MITSuME images with $9\times1$-minute exposure 
or we judge the detection in the subtracted KWFC images are reliable, 
we quickly trigger 
multi-facility, multi-mode follow-up observations under the umbrella of 
the KISS collaboration. 
The collaborative follow-up observations so far include,
in order of telescope aperture,
those with 
50~cm-MITSuME telescope located at Okayama Astrophysical Observatory \citep{yanagisawa2010}, 
the 70-cm telescope of Sternberg Astronomical Institute,  
KPNO 0.9-m, 
Lulin One-meter Telescope (LOT), 
Atacama Near-Infrared Camera~(ANIR; \cite{motohara2010,konishi2014}) on the 1-m miniTAO telescope \citep{sako2008,minezaki2010}, 
HOWPol \citep{kawabata2008} and HONIR \citep{sakimoto2012} on the 1.5-m Kanata telescope, 
the Kyoto Okayama Optical Low-dispersion Spectrograph (KOOLS; \cite{yoshida2005}) on the Okayama-188cm telescope, 
the Himalayan Faint Object Spectrograph (HFOSC) on the 2-m Himalayan Chandra Telescope (HCT), 
the Wide Field CCD Camera (WFCCD) on the Las Campanas 2.5-m du Pont telescope, 
the Andalucia Faint Object Spectrograph and Camera (ALFOSC) on the 2.5-m Nordic Optical Telescope (NOT), 
the Device Optimized for the LOw RESolution (DOLORES) on the 3.58-m Telescopio Nazionale Galileo (TNG; \cite{barbieri1997}), 
the FoldedPort Infrared Echellette (FIRE) spectrograph on the 6.5-m Magellan Baade Telescope, 
and Faint Object Camera And Spectrograph (FOCAS; \cite{kashikawa2002}) 
on the 8.2-m Subaru telescope. 

\subsection{Number Estimate of Supernova Shock Breakout Detections and Theoretical Predictions}
\label{sec:numberestimate}
Based on the theoretical model of a shock breakout which succeeded in 
reproducing the UV light curve of the shock breakout and optical plateau multi-band 
light curves for SNLS-04D2dc \citep{tominaga2009}, 
we optimize the KISS survey parameters as done in \citet{tominaga2011} (see Equation 3)
where they describe the expected performance of 
future large surveys with larger aperture telescopes. 
Whereas \citet{tominaga2011} adopt the cosmic star formation history to derive expected 
numbers of shock breakouts up to $z\sim3$, 
we first estimate the number of detections 
assuming a constant star formation rate density with redshift 
to evaluate how much observing time 
for a given star formation rate (i.e., directly translated to core-collapse SN rate) we need to detect a shock breakout. 
We here assume a detection limit of $g_{\rm{sub}}=19.8$~mag 
in our subtracted images, the typical depth
as shown in \S~\ref{sec:transientdetection}, 
for calculating the effective volume. 
We simulate light curves with various observation strategies; 
different cadence (time interval between exposures for a given field $t_{\rm{int}}$, 0.5-2~hours);  
different numbers of exposures per night $N_{\rm{visit}}$ (3-5), 
(i.e., different number of the fields $N_{\rm{field}}$ within a fixed observing time). 
In this paper, the detection of shock breakouts is defined 
so that the object is detected 
in 1 or more epochs around the shock breakout peak 
($-0.1<t<0.1$~days 
where $t$ is measured relative to the peak)
and 
in 3 or more epochs over wider time range including the plateau phases. 
As a result, it is shown that the number of detections per unit star formation rate is 
20-50\% larger 
for longer time interval $t_{\rm{int}}$ 
and the total number of detection per night is only 20\% larger 
in the $N_{\rm{visit}}=3$ case than that in the $N_{\rm{visit}}=5$ case. 
We also estimate the observable redshift for each different survey strategy 
and find that 
about half of the detections are located at $z<0.02$ and 90\% of them are at $z<0.03$ 
in every observation strategy. 

\begin{figure}
  \begin{center}
	 \includegraphics[angle=270,width=80mm]{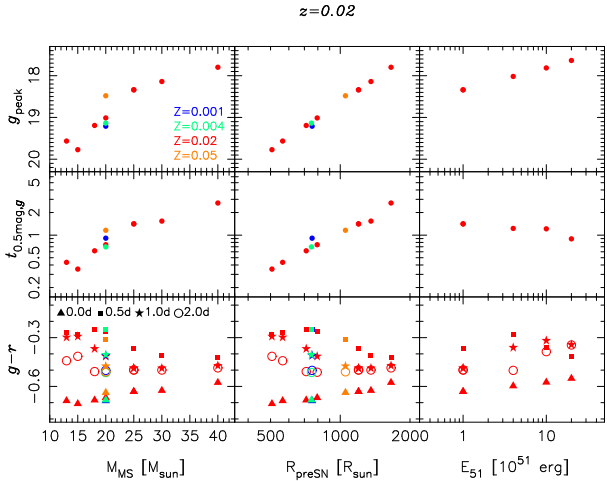}
  \end{center}
  \caption{
The dependency of observational quantities, 
peak $g$-band magnitude $m_{g,\rm{peak}}$, 
hours for $0.5$~mag decline from the peak in $g$-band $t_{g,\rm{0.5mag}}$, 
and $g-r$ at $z=0.02$ 
for different model parameters; main-sequence masses $M_{\odot}$, presupernova radius $R_{\rm{preSN}}$, 
and explosion energies $E_{51}$ from left to right. 
Main-sequence mass $M_{\odot}$ in the three right panels are 25~$M_{\odot}$. 
 Different colors indicate different metallicity; $Z=0.001, 0.004, 0.02$, and 0.05 in blue, green, red, and orange, respectively. 
 Different symbols (filled triangles, boxes, stars, and open circles) also indicate different epochs; $t=0.0, 0.5, 1.0$, and $2.0$~days 
 from the time with the bluest $g-r$ color, slightly after the $g$-band peak. 
 }
\label{fig:physicalparam}
\end{figure}

It is important to enhance not only the number of the detected objects but also
the number of the detected exposures for individual objects 
to identify the shock breakouts and characterize the progenitor stars.
The top two rows of Figure~\ref{fig:physicalparam} show the dependencies
of observational quantities on the theoretical model parameters; 
a brighter peak luminosity and slower decline rate of the shock
breakout result from a progenitor star with a larger presupernova radius, i.e., higher main sequence mass. 
Although core-collapse SNe usually explode in dusty environment and are obscured by dust, 
some of the observational properties (e.g., decline rate) are free from dust extinction. 
Therefore, we conclude that 
the survey parameters as described in \S\ref{sec:observingsetup}, i.e.,
1-hour cadence ($t_{\rm{int}}=1$~hour) and 5 visits per night ($N_{\rm{visit}}=5$)
using a single $g$-band, which corresponds to 20 fields ($N_{\rm{field}}=20$),
would be the optimum strategy.
In this survey parameter set, the expected number of shock breakout detections 
is an order of 1 during the 3-year project term 
considering all the factors described above and 
the typical SFR per KWFC FoV is 20~$M_\odot$ yr$^{-1}$ at $z<0.03$.

\begin{figure}
  \begin{center}
	 \includegraphics[angle=270,width=80mm]{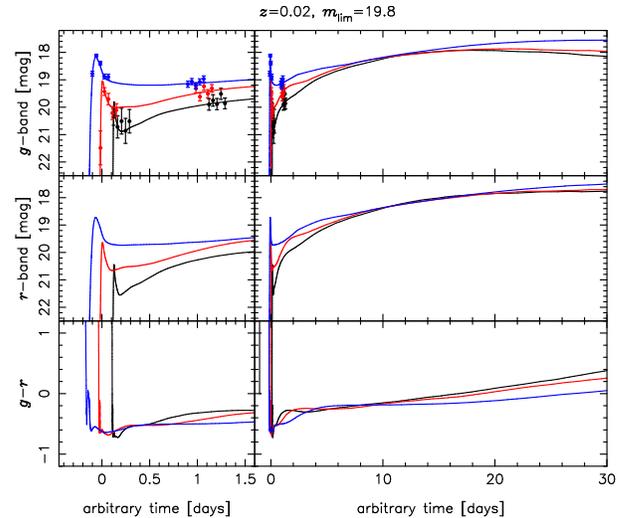}
  \end{center}
  \caption{
Simulated $g$-band light curves for $M=15$ (black), $20$ (red), $30~M_\odot$ (blue) 
progenitor stars at $z=0.02$ (at a distance of $\sim85$~Mpc). 
Time in $x$-axis is arbitrarily shifted for ease of comparison. 
Photometric errors are calculated by assuming a limiting magnitude ($5\sigma$) of $19.8$~mag. 
The left panels are magnified views of the right panels around the shock breakout phases. 
}
\label{fig:simulatedlc}
\end{figure}

Figure~\ref{fig:simulatedlc} shows simulated KWFC $g$-band light curves
for shock breakouts of stars with $M=15,~20,~{\rm and}~30M_\odot$ at a
distance of $d=85$~Mpc ($z\sim0.02$) by
assuming the typical depth ($g_{\rm{sub}}=19.8$~mag, $5\sigma$) and 
observing strategy of our survey. 
The figure demonstrates that the decline of the light curve after the
shock breakout 
can be well characterized by our survey strategy.
While the light curves are almost constant on the second day, the SN brighten with time again 
and the luminosities in $g$-band peak at $20-30$~days after the explosion
(Figure~\ref{fig:simulatedlc}). The $g$-band and $r$-band magnitudes
at the plateau phase are expected to be as bright as $<18$~mag at $z=0.02$. 
The plateau duration and the
brightness and photospheric velocity at the plateau phase also depend on
the presupernova radius, envelope mass, and explosion energy
\citep{eastman1994} and have been frequently used to constrain the
explosion properties \citep{utrobin2009}. Therefore, in order to confirm
the detection of the shock breakout and verify the constraints derived
from the observational properties of the shock breakout, it is important
to continuously observe the same fields 
over several months. 

Further constraints on physical properties of shock breakouts discovered by KISS 
would be achieved with the quick follow-up observations described in \S\ref{sec:followupobs}. 
For example, the change in time of observed colors such as $g-r$ in the shock breakout phase 
is also sensitive to the progenitor mass 
as shown in the bottom row of Figure~\ref{fig:physicalparam}. 
The color evolution
indicates
the photospheric temperature and constrains the
evolution of the photospheric radius, i.e., the shock velocity. This
additional information determines the presupernova radius and the
explosion energy independently. Furthermore, if spectroscopic
observations are performed, growing metal absorption lines due to the
decrease of the photospheric temperature will be observed
\citep{gezari2008}. The spectral evolution provides another clue to  
properties of shock breakouts and supernovae, e.g., presupernova radius,
circumstellar material 
structure, and mass loss at the last stage of stellar evolution. 
It also reveals the structure of a radiation-mediated shock, 
in which radiation and matter are marginally
coupled. This can test radiation hydrodynamics theories at high
temperature, e.g., how the absorptive and scattering opacities
contribute to the total opacity \citep{blinnikov2000}.

\begin{figure}
  \begin{center}
    \includegraphics[angle=270,width=80mm]{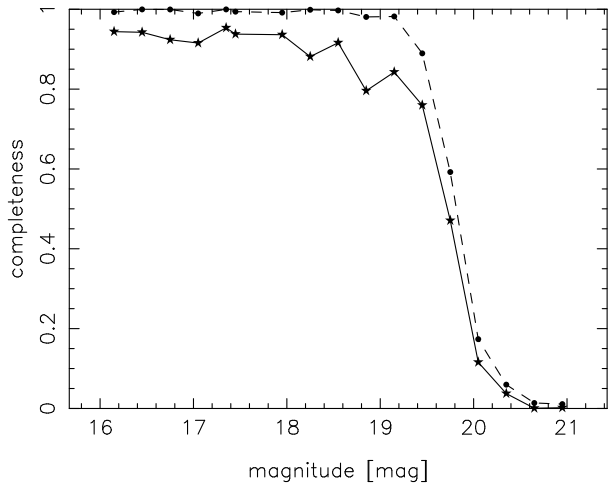}
  \end{center}
  \caption{
Detection completeness as a function of magnitude
estimated by embedding artifical sources in the KWFC images with 5$\sigma$ detection limit of 20.5~mag.
Dashed and solid lines show the completeness before and after screening (based on position, FWHM, and elongation).
The missed fraction of real sources is about 10 \% as we adopt relatively stringent thresholds 
to achieve realtime detection of fast transients and to avoid too many spurious detection.
}
\label{fig:completesub}
\end{figure}

\section{Data Reduction and Transient Detection}
\label{sec:datareductiontransientdetection}
Here we summarize the data analysis procedure optimized for the KISS/KWFC data.
Shock breakout is very rare and
the time scale is as short as hours.
It is essential to find good candidates
with high completeness and a low contamination rate
from a huge amount of data as soon as possible,
ideally automatically.

\subsection{Automatic Realtime Data Reduction}
\label{sec:datareduction}
Data reduction starts automatically just after KWFC data are acquired.
First, a standard reduction which includes overscan subtraction, 
overscan region trimming, bias subtraction,
flat-fielding, and background subtraction are performed for each image. 
The bias and flat-field data used here are prepared before the night
observations start.
Then, astrometric solutions are also automatically 
obtained with the USNO-B1.0 catalogue adopted 
as the reference list using the Optimistic Pattern Matching (OPM) algorithm \citep{tabur2007} 
implemented by one of the authors (NM).
Typically 100-200 stars in the USNO-B1.0 catalogue are identified 
within each chip in our KISS observations ($g$-band, 180~sec exposure, not in the Galactic Plane) 
and used to obtain a linear transformation from the pixel coordinate to the celestial coordinate.
The RMS residual is about 0.3~arcsec in both RA and Dec directions. 
Finally, calculations of the zeropoint magnitude and limiting magnitude 
(50\% detection completeness magnitude) 
are done using the SDSS photometric catalog.
These procedures are performed for each chip individually.
It typically takes about 5~minutes to finish these procedures
for each exposure (all 8 chips).

\subsection{Realtime Transient Detection}
\label{sec:transientdetection}

SDSS archival $g$-band images taken several years before the KWFC observations
are used as reference images for image subtraction.
The SDSS images are always deeper 
($g\sim22.2$~mag, 95\% detection repeatability for point sources)
and sharper ($\sim1.4$~arcsec PSF width in $r$-band 
according to the SDSS DR8 website\footnote{https://www.sdss3.org/dr8/scope.php})
with finer spatial sampling of 0.396~arcsec pixel$^{-1}$,
compared to the KWFC images ($20-21$~mag limit,
$3.3-5.3$~arcsec PSF FWHM, and 0.946~arcsec pixel$^{-1}$).
Transmission curves of the $g$-bands of these two datasets (SDSS and KWFC)
are slightly different 
but this difference does not significantly affect 
the quality of image subtraction,
although we need to take care of objects with extreme colors, 
i.e., steep spectrum in the $g$-bands.

Our transient detection pipeline performs following procedures.
First, cosmic rays are removed from the reduced KWFC data using 
the publicly available code,
{\it L.A.Cosmic}\footnote{http://obswww.unige.ch/\~{}tewes/cosmics\_dot\_py/}
developed by \citet{vandokkum2001}.
The SDSS images prepared in advance are transformed to the KWFC images
with {\it wcsremap}\footnote{http://www.astro.washington.edu/users/becker/v2.0/wcsremap.html},
and then subtracted from the KWFC image using
{\it hotpants}\footnote{http://www.astro.washington.edu/users/becker/v2.0/hotpants.html}.
In the subtracted images, we search only for positive residuals
with $>5\sigma$ detections using the {\it SExtractor} 
software \citep{bertin1996}.
In this way, we obtain transient candidate catalogs about 10~minutes after the data acquisition.

Each exposure typically gives about 1000-1500 positive detection
in the subtracted image.
However, as in other transient surveys using image subtraction techniques,
non-astrophysical sources always 
overwhelm the real sources 
(\cite{bloom2012}; \cite{brink2013}). 
Non-astrophysical sources include cosmic rays that are not removed by
{\it L.A.Cosmic}, extended features around very bright saturated stars,
and bad subtractions due to a misalignment of the images.
To screen out these false detection,
we first exclude the sources around bright, saturated stars
and at the edges of the images.
Next, we apply criteria based on the FWHM and elongation of the detected sources.
Using these screening procedures, 
we reduce the number of candidates down to 10-50 per exposure.
Even after the screening, there are still false detections,
but the numbers of non-astrophysical and astrophysical sources 
are roughly comparable.
The false positive rate (the fraction of non-astrophysical sources
that are judged to be real objects) of our pipeline is about 1-5 \%.

We estimate the detection completeness of our pipeline 
by embedding artificial point sources in the KWFC images
with {\it IRAF/artdata} 
and detecting them using the transient detection pipeline
in exactly the same way as for the original images.
Figure \ref{fig:completesub} shows a typical example of detection completeness as a function of magnitude.
Detection completeness is about 0.9 after screening
even for the bright sources,
i.e., the fraction of real sources that are missed by our pipeline
is about 10\%.
There is an inevitable 
trade-off between the false positive rate
and the missed fraction (\cite{bloom2012}; \cite{brink2013}), 
and we adopt relatively stringent thresholds to achieve realtime
detection of fast transients while avoiding too many spurious detections.
For the same reason, the detection limit in the subtracted images is 
typically 0.5-1.0~mag shallower than that of the original images.

After running the transient detection pipeline, 
we list all the remaining candidates on a web page for visual inspection
to further remove non-astrophysical sources and classify real objects.
Real objects include not only SN but also 
variable stars, AGN, and moving objects such as asteroids.
For the classification,
we cross-match the detected object catalog with
quasar and AGN catalogs \citep{veroncetty2010,paris2014},
X-ray sources from ROSAT All-Sky catalogs (Bright Source Catalog; BSC; \cite{voges1999};
Faint Source Catalog; FSC), 
the XMM serendipitous catalog \citep{watson2009},
Chandra X-ray sources, 
the Minor Planet Checker ({\it MPChecker}\footnote{http://scully.cfa.harvard.edu/cgi-bin/checkmp.cgi}),
and variable stars from SIMBAD.
For rapid and efficient visual screening,
we ask volunteer amateur astronomers in Japan
to check whether the candidates are real astronomical sources or not and SNe or not.
When we judge that a candidate is a real astronomical source and
a good candidate for a newly discovered supernova,
we trigger follow-up observations as described in \S\ref{sec:followupobs}.
We note that all the SNe reported in CBET by other groups within our survey area
were not missed by our reduction system.

\section{Initial Results}
\label{sec:initialresults}

We discovered about 80~supernova candidates up to the end of May 2014. 
Sixteen SNe among them were spectroscopically identified within the KISS collaboration and reported to CBET. 
In this paper, we do not describe any details of SN~2014bk \citep{morokuma2014b,stritzinger2014}, 
a Type~Ibn SN, which will be discussed in our forthcoming paper \citep{morokuma2014c}. 
Table~\ref{tab:kisssnlist} summarizes the properties of fifteen SNe except for SN~2014bk 
which include 8~core-collapse SNe and 7~SNe~Ia. 
Table~\ref{tab:specobs} shows a summary of spectroscopic observations 
for identification.

\begin{figure}
  \begin{center}
    \includegraphics[width=85mm]{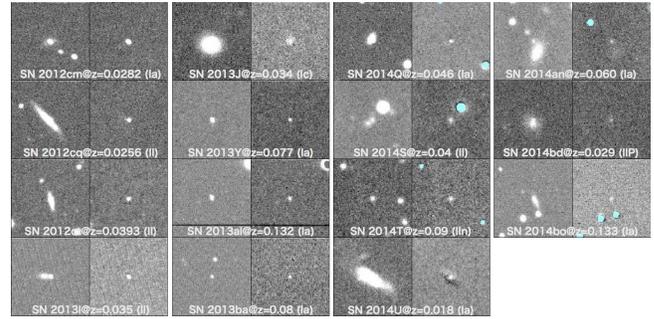} 
  \end{center}
  \caption{
Discovery images of the KISS SNe which were reported to CBET except for SN~2014bk. 
The image sizes are $2$~arcmin$\times2$~arcmin. 
North is up and east is left. 
}
\label{fig:discimages}
\end{figure}

\begin{figure}
  \begin{center}
    \includegraphics[angle=270,width=80mm]{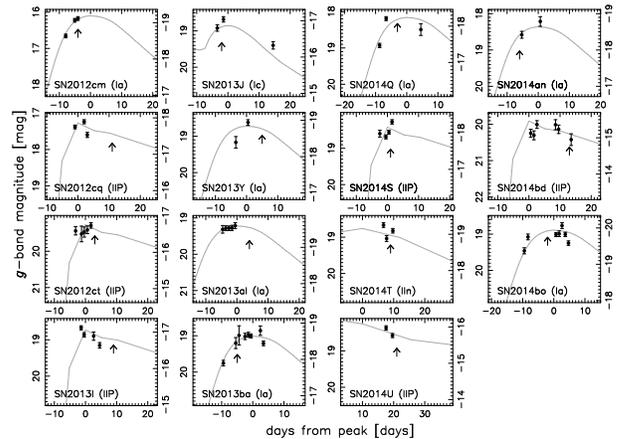}
  \end{center}
  \caption{
Kiso/KWFC $g$-band light curves for the CBET-reported KISS SNe except for SN~2014bk. 
Nugent's light curve templates \citep{nugent2002} except for SN~2013J,
and Drout's SN~Ic template \citep{drout2011} for SN~2013J, are overlaid in gray lines. 
Arrows indicate the dates of the spectroscopic observations. 
The right-hand $y$-axis indicates the absolute magnitude in $g$-band without K-corrections. 
}
\label{fig:snlc}
\end{figure}

\begin{figure}
  \begin{center}
    \includegraphics[angle=270,width=80mm]{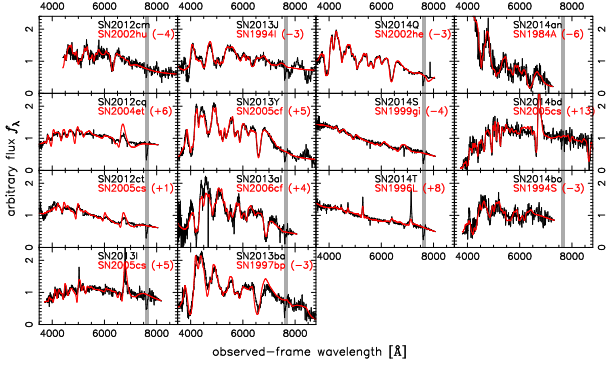}
  \end{center}
  \caption{
Identification spectra for the CBET-reported KISS SNe except for SN~2014U 
(shown in Figure~\ref{fig:snspec2}) and SN~2014bk. 
The flux density ($f_\lambda$) scale has arbitrary units. 
Observed and template spectra are shown in black and red, respectively. 
SN spectra of the best-fit templates are also shown in red characters with the 
estimated phase (days relative to max light) in parentheses. 
Telluric absorption wavelength regions around 7,600\AA\ are gray-shaded.
 }
\label{fig:snspec}
\end{figure}

\begin{figure}
  \begin{center}
    \includegraphics[angle=0,width=80mm]{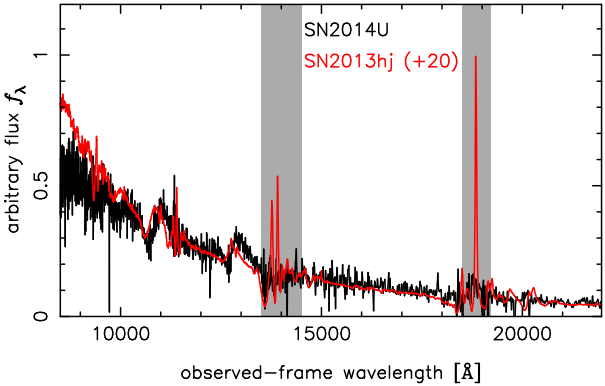}
  \end{center}
  \caption{
Identification near-infrared spectra for SN~2014U. 
Flux density ($f_\lambda$) scale is in arbitrary unit. 
Observed and template (SN~2013hj at $t=+20$ days) spectra are shown in black and red, respectively. 
Telluric absorption wavelength regions are gray-shaded.
}
\label{fig:snspec2}
\end{figure}

Figure~\ref{fig:discimages} and Figure~\ref{fig:snlc} show the discovery KWFC images and 
KWFC $g$-band light curves of 
the fifteen SNe. 
The light curve templates overplotted on the KWFC light curves for Figure~\ref{fig:snlc} 
are Nugent's templates \citep{nugent2002} 
except for a Type~Ic SN (SN~2013J) for which 
a $V$-band light curve template is derived 
from \citet{drout2011}\footnote{https://www.cfa.harvard.edu/\~{}mdrout/SN\_Templates.html}. 
We do not fit the light curves with these templates, 
and time shifts of a few days from the spectroscopic epochs and 
peak magnitude shifts from the typical peak magnitudes are allowed. 
The absolute maximum magnitudes are also listed in Table~\ref{tab:kisssnlist}, 
which are obtained assuming the luminosity changes of the light curve templates. 
All the SN spectra except for that of SN~2014U are cross-correlated with a library of supernova spectra 
with the SNID code \citep{blondin2007} and shown in Figure \ref{fig:snspec} (also except for SN~2014bk). 
The SN~2014U near-infrared spectrum 
is shown in Figure \ref{fig:snspec2} with the spectrum of SN~2013hj at $t=21$~days overplotted. 
The spectroscopic phase $t_{\rm{spec}}$ is first calculated from the SNID fitting results 
and then discovery epochs $t_{\rm{disc}}$ are derived from spectroscopic phase $t_{\rm{spec}}$ and 
time difference between the discovery and spectroscopic observations. 
We provide observational results for the individual SNe in the appendix. 
Data of these light curves and identification spectra for the fifteen SNe are available  
in our KISS website, http://www.ioa.s.u-tokyo.ac.jp/kisohp/KISS/datarelease\_en.html. 

\begin{figure}
  \begin{center}
    \includegraphics[angle=270,width=80mm]{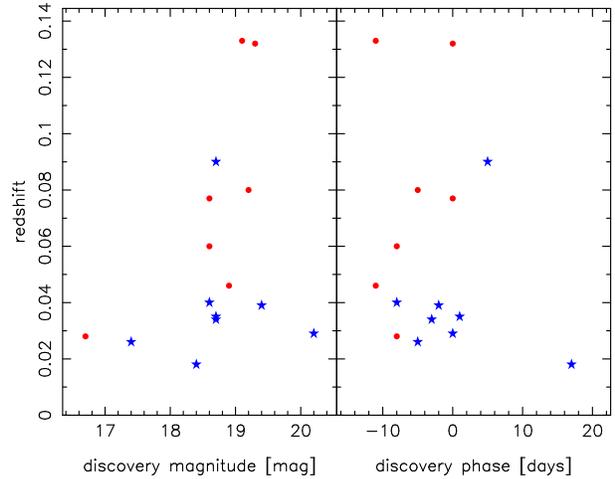}
  \end{center}
  \caption{
Distributions of the fifteen CBET-reported SNe except for SN~2014bk 
in discovery magnitude versus redshift in the left panel 
and discovery phase versus redshift in the right panel. 
SNe~Ia and core-collapse SNe are shown in red circles and blue stars, respectively. 
}
\label{fig:snstat1}
\end{figure}

Apparent $g$-band discovery magnitudes and discovery phases $t_{\rm{disc}}$ 
as functions of redshift are shown in the left and right panels of Figures \ref{fig:snstat1}, respectively. 
Discoveries of most of the SNe are at relatively early phases, 
i.e., before or around maximum light,
although we have not detected shock breakouts 
because SNe discovered by KISS are as distant as $z>0.02$ 
and only the bright parts of the SN light curves are detected. 
This is not unexpected because the survey volume at higher redshifts is larger. 
Another reason is that our observations are 
conducted only for $\sim10$~continuous nights around new moon
and some time was lost due to bad weather. 
Typically our observation limits correspond to Type~Ia SNe at $z\sim0.1$ and 
core-collapse SNe at $z\sim0.04$. 

The short time scale of $\sim1$~hour investigated in the KISS has been a new parameter space, 
in which transient phenomena have not been explored enough \citep{ivezic2008,kasliwal2011}. 
This new survey would have potential to provide serendipitous detection of rapid variability as a by-product. 
For example, we actually detected a rapid flare of a radio-loud narrow-line type-1 AGN 
at $z=0.840$ \citep{tanaka2014f}, which may be similar to intranight variability of 
$\gamma$-ray-loud narrow-line Seyfert~1 galaxies \citep{maune2013,itoh2013}. 

\section{Summary}
\label{sec:summary}

We describe a new SN survey, KISS, optimized for 
detecting SN shock breakouts using high-cadence wide-field imaging observations 
with the KWFC on the 1.05-m Kiso Schmidt telescope. 
Based on theoretical models, 
we simulate observed light curves of shock breakouts and those in the plateau phases 
and find that the best strategy to detect, identify, and characterize shock breakouts 
is observations 
in a single bandpass, $g$,
with a 1-hour cadence ($t_{\rm{int}}$), 5 visits per night ($N_{\rm{visit}}$), 
corresponding to 20 fields ($N_{\rm{field}}$).
The expected number of shock breakout detections is of order 1 during the 3-year KISS project duration. 
Multi-band photometry obtained with quick follow-up observations 
would provide detailed information on the progenitor stars. 
We have developed quick automatic data reduction and image subtraction systems, 
including human visual screening by us and our volunteers, 
which would provide candidate catalogs about 15~minutes after the observations.
We also summarize our discovery, follow-up observations and spectrum identification results 
for the sixteen CBET-reported SNe including 9 core-collapse SNe and 7 SNe~Ia. 

\bigskip

We thank the anonymous referee for providing helpful comments and suggestions. 
This work has been partly supported by the Grants-in-Aid of the 
Ministry of Education, Science, Culture, and Sport [23740143, 25800103~(TM), 23740157~(NT), 24740117, 25103515~(MT)], 
by the RFBR-JSPS bilateral program (RFBR grant No. 13-02-92119), 
the National Science Foundation under Grant No.~AST-1008343, 
FUND::::INAF PRIN 2011 and PRIN MIUR 2010/2011, 
and Optical \& Near-Infrared Astronomy Inter-University Cooperation Program, supported by the MEXT of Japan. 

This work is also based on data obtained 
with the Nordic Optical Telescope, operated by the Nordic Optical Telescope Scientific Association at the Observatorio del Roque de los Muchachos, La Palma, Spain, of the Instituto de Astrofisica de Canarias, 
 with the Italian Telescopio Nazionale
Galileo (TNG) operated on the island of La Palma by the Fundaci\'{o}n Galileo Galilei of the INAF (Istituto Nazionale di Astrofisica) at the
Spanish Observatorio del Roque de los Muchachos of the Instituto de Astrofisica de Canarias, 
with ALFOSC, which is provided by the Instituto de Astrofisica de Andalucia (IAA) 
under a joint agreement with the University of Copenhagen and NOTSA, 
with the 6.5-m Magellan~I Baade telescope, located at Las Campanas Observatory, Chile. 
and with Subaru Telescope, which is operated by the National Astronomical Observatory of Japan. 
SDSS-III is managed by the Astrophysical Research Consortium for the Participating Institutions of the SDSS-III Collaboration including the University of Arizona, the Brazilian Participation Group, Brookhaven National Laboratory, Carnegie Mellon University, University of Florida, the French Participation Group, the German Participation Group, Harvard University, the Instituto de Astrofisica de Canarias, the Michigan State/Notre Dame/JINA Participation Group, Johns Hopkins University, Lawrence Berkeley National Laboratory, Max Planck Institute for Astrophysics, Max Planck Institute for Extraterrestrial Physics, New Mexico State University, New York University, Ohio State University, Pennsylvania State University, University of Portsmouth, Princeton University, the Spanish Participation Group, University of Tokyo, University of Utah, Vanderbilt University, University of Virginia, University of Washington, and Yale University. 
This research has made use of data obtained from the Chandra Source Catalog, provided by the Chandra X-ray Center (CXC) as part of the Chandra Data Archive. 
We acknowledge all the members of the KISS amateur team who use their time 
and experience to discriminate  
real astronomical objects from spurious detections in the subtracted images. 
As of May 2014, 18~people and 2~groups, 
including 
Satoru Fukuda, 
Hiroki Iida, 
Katsuhiko Mameta, 
Jun Shimizu, 
Koichi Takahashi, 
Masanori Takeishi, 
Shingo Tanaka, 
Naoto Tatsumi, 
Hirofumi Ueda, and 
Keiichi Yoshida, 
have joined the screening activity. 

\appendix
\section{SNe Discovered and Reported to CBET by KISS}
\label{sec:eachsn}

We note that apparent magnitudes shown in this appendix are 
the SN brightness measured in the daily-stacked KWFC $g$-band images, 
which are different from the brightness in our CBET reports. 

\begin{itemize}
%
%
%
\item	SN~2012cm\\
SN~2012cm ($g=16.8$) was discovered 
on May 13.6, 2012 at (09h10m05s.45, +20d12'45.1''). 
SN~2012cm is located 1.2'' east and 0.5'' south of the host galaxy, SDSS~J091005.37+201245.6, at $z_{\rm{SDSS}}=0.0282$. 
Nothing is detected at this position on SDSS and KWFC reference or subtracted images taken on and before Apr. 28, 2012.
A spectrum taken with HOWPol on the 1.5-m Kanata telescope on May 17.5, 2012
shows clear Si II, S II, and Fe III features. 
Spectral fitting with the SNID code indicates that the spectrum is 
most similar to 4~days before maximum of a slightly overluminous Type~Ia SN, SN~2002hu \citep{sahu2006a}. 

\item	SN~2012cq\\
SN~2012cq ($g=18.2$)  was discovered on May 13.5, 2012  at (09h08m05s.46, +20d30'12''.5). 
SN~2012cq is located 4''.8 west and 0''.1 south of the host galaxy, 
SDSS~J090805.78+203012.6 (UGC~4792; \cite{giovanelli1997}) at $z_{\rm{SDSS}}=0.0256$. 
Nothing was detected at this position on the SDSS and KWFC reference or subtracted images taken on and before Apr. 28, 2012.
SN~2012cq was confirmed after discovery with the Lulin 1-m telescope
and the HOWPol on the 1.5-m Kanata telescope. 
$g$-band magnitude was measured to be 18.1 on May 14.5 and 16.5, 2012.  
A spectrum was taken with DOLORES on the 3.5-m TNG on May 24.9, 2012. 
The spectrum shows a series of hydrogen Balmer features 
and is similar to that of a Type~IIP SN, SN~2004et 
\citep{sahu2006b}, at 6 days after the maximum according to the SNID code. 

\item	SN~2012ct\\
SN~2012ct ($g=18.0$) was discovered SN~2012ct on May 22.5, 2012 at (16h32m13s.92, +38d39'25''.1). 
SN~2012ct is located 2''.7 east and 5''.1 north of the host galaxy, 
SDSS~J163213.74+383920.1, at $z_{\rm{SDSS}}=0.0393$.
Nothing was detected at this position on reference images taken on and before May 12, 2012. 
A spectrum was taken with DOLORES on the 3.5-m TNG on May 24.9, 2012. 
The spectrum shows a series of hydrogen Balmer features indicating the SN~2012ct is a Type~II SN. 
According to spectral fitting with the SNID, the SN~2012ct spectrum is most similar to 
a moderately subluminous Type~IIP SN appeared in M51, 
SN~2005cs \citep{pastorello2009} around the maximum light. 

\item	SN~2013I\\
SN~2013I ($g=18.6$) was discovered on Jan. 11.4, 2013 at (02h49m42s.17, +0d45''35''.7). 
SN~2013I is located 8''.8 west and 0''.3 south of the host galaxy, SDSS~J024942.76+004535.4 at $z_{\rm{SDSS}}=0.035$. 
This object was also marginally detected at $g\sim20$ in a previous image
taken on Nov. 6.6, 2012, but nothing was seen at this position in an image
taken on Oct. 21.8, 2012 or on the SDSS image. 
SN~2013I was also detected ($R=17.8$) with HOWPol on the 1.5-m Kanata telescope on Jan. 12.4, 2013. 
We note that SN~2013I was independently discovered by the Dark Energy Supernova Survey (DES12S1a; \cite{brown2013}). 
A spectrum was taken ALFOSC on the 2.5-m NOT on Jan. 15.9, 2013. 
The spectrum shows multi-component H$\alpha$ and H$\beta$ emission features, 
superposed on a continuum containing a number of broad absorption features. 
According to spectral fitting with the SNID code, 
the SN~2013I spectrum resembles a spectrum of a Type IIP SN, SN~2005cs, 
around 5 days after the maximum as well as the SN~2012ct spectrum. 

\item	SN~2013J\\
SN~2013J ($g=18.8$) was discovered on Jan. 19.8, 2013 at (11h12m50s.30, +28d04'19''.7). 
SN~2013J is located 1''.5 west and 5''.5 north of the host galaxy (KUG~1110+283 or SDSS~J111250.41+280414.2 
at $z_{\rm{NED}}=0.034$) \citep{mahdavi2004}. 
This object was also marginally detected at $g\sim20$ 
in our previous image taken on Jan. 15.8, 2013, but nothing is seen at this position in an image taken
on and before Jan. 12.8, 2013 or in the SDSS image. 
The light curve matches that of the Type~Ic template in \cite{drout2011} better than that in \cite{nugent2002}, 
which is not surprising as the Drout's sample is larger and more recent than Nugent's sample.
A spectrum was taken with ALFOSC on the 2.5-m NOT on Jan. 21.2, 2013. 
According to the spectral fitting with the SNID code, 
the spectrum is most similar to the maximum-light ($t\sim-2$~days) 
spectrum of a fast-fading Type~Ic SN, SN~1994I \citep{fillipenko1995}. 

\item	SN~2013Y\\
SN~2013Y ($g=18.7$) was discovered on Feb. 6.7, 2013  at (12h09m39s.70, +16d12'14''.3). 
The SN is located 1''.2 east and 2''.1 north of the host galaxy, SDSS~J120939.62+161212.2 at $z_{\rm{SDSS}}=0.077$. 
Nothing is seen at this position in the SDSS image. 
This object was confirmed with the Swope 1-m telescope at Las Campanas Observatory.
A spectrum was taken with ALFOSC on the 2.5-m NOT on Feb. 11.2, 2013. 
Spectral fitting with the SNID code indicates that the spectrum is most similar to that of a very normal Type~Ia SN, SN~2005cf 
(\cite{garavini2007}; \cite{wang2009}) at 5~days past maximum light. 

\item	SN~2013al\\
SN~2013al ($g=19.2$) was discovered on Mar. 3.4, 2013 at (11h14m54s.07, +29d35'06''.0). 
SN~2013al is located 3.0'' south of a possible host galaxy, SDSS~J111454.06+293508.6. 
Nothing is seen at this position in an image taken on Feb. 6.7, 2013. 
This object is also detected with
the WIYN 0.9-m telescope at Kitt Peak at magnitudes $V=19.4$, $R=19.1$, and $I=19.3$ on Mar. 5.2, 2013. 
A spectrum was taken with DOLORES on the 3.58-m TNG on Mar. 7.1, 2013.  
According to spectral fitting with the SNID, 
the best match to the spectrum of SN~2013al is the Type~Ia SN, SN~2006cf 
at $t=+2$~days.  
The redshift of the host galaxy is derived from the same spectrum data and found to be $z_{\rm{host}}=0.1321$ 
from its [O II], H$\beta$, [O III], and H$\alpha$ emission lines of the host galaxy. 

\item	SN~2013ba\\
SN~2013ba ($g=19.9$) was discovered on Apr. 4.7, 2013 at (13h52m56s.63, +21d56'21''.7). 
SN~2013ba is located 0''.7 east and 0''.8 north of a possible host galaxy, SDSS~J135256.58+215620.9. 
Nothing is seen at this position in images taken on Mar. 10.8, 2013. 
The SN was confirmed with the Swope 1-m telescope at Las Campanas Observatory. 
A spectrum was taken with ALFOSC on the NOT on Apr. 7.2, 2013. 
According to spectral fitting with the SNID code, the SN~2013ba spectrum is most similar to that of a normal Type~Ia SN, 
SN~1997bp \citep{altavilla2004} a few days before maximum light. 
The redshift of $z_{\rm{SN}}=0.08$ is derived from the SN spectrum fitting.

\item	SN~2014Q\\
SN~2014Q ($g=19.2$) was discovered on Jan. 29.4, 2014 at (08h18m50s.20, +57d06'02''.9). 
SN~2014Q is located 2''.4 west and 2''.7 south of the host galaxy, 
SDSS~J081850.49+570605.5; $z_{\rm{SDSS}}=0.046$. 
The SN was confirmed with the optical three-color CCD cameras on the MITSuME 50-cm
telescope of the Akeno Observatory, on Feb. 1.5, 2014. 
The SN was also marginally detected in a $g$-band KWFC image on Jan. 27.4, 2014, 
but nothing is seen at this position in an image taken on Jan. 6.5, 2014 or in the previous SDSS image. 
A spectrum was taken with DOLORES on the 3.58-m TNG on February 6.0, 2014. 
The spectrum of SN~2014Q is found to be most similar to that of a normal Type~Ia SN, 
SN~2002he around $3$ days before maximum. 
Optical spectra of this object were also obtained with KOOLS on the Okayama 188-cm telescope 
on Jan. 30.8 and 31.8, 2013 and the continuum was significantly detected.

\item	SN~2014S\\
SN~2014S ($g=18.8$) was discovered on Feb. 21.5, 2014 at (10h40m24s.98, Decl. = +53d57'58''.6). 
The SN is located 6''.6 west and 5''.4 north of the presumed host galaxy, SDSS~J104025.73+535753.2. 
Nothing is detected at this position in an image taken on Jan. 31.6, 2014. 
SN~2014S was also detected with HONIR on the Kanata 1.5-m telescope 
with the following magnitudes measured, $J=18.7$ on Feb 22.6, 2014, $R_c = 18.4$ on Feb 22.6, 2014. 
A spectrum was taken with DOLORES on the 3.58-m TNG on Feb. 25.1, 2014. 
The spectrum shows a strong blue continuum with faint emission lines of
H$\gamma$, H$\delta$, and He~I, corresponding to a redshift of $z_{\rm{SN}}=0.04$. 
Using the SNID code, the best matches to the
spectrum of SN~2014S are to a Type~IIP SN, SN~1999gi \citep{leonard2002} 4~days before the maximum.  
Due to the faint emission lines, distinguishing between a Type~IIP or a Type~IIn event is difficult.

\item	SN~2014T\\
SN~2014T ($g=18.7$) was discovered on Feb. 22.7, 2014  at a position (14h36m04s.98, +02d20'34''.2). 
SN~2014T is located 1''.5 east and 1''.6 north of the presumed host galaxy, SDSS~J143604.90+022032.6. 
Nothing is seen at this position in an image taken on May 9.7, 2013. 
A spectrum was taken with DOLORES on the 3.58-m TNG on Feb. 25.2, 2014. 
The spectrum shows a blue continuum with strong emission lines of H$\alpha$,
H$\beta$, H$\gamma$, and H$\delta$ and also He~I at a redshift of $z_{\rm{SN}}=0.09$. 
Using the SNID code, many good matches to the SN~2014T spectrum are spectra of galaxies and AGN
but the best match is to the spectrum of a Type~IIn SN, SN~1996L at $8$ days past explosion, based on visual inspection. 

\item	SN~2014U\\
SN~2014U ($g=18.9$) was discovered on Feb. 23.5, 2014  at a position (11h44m52s.16, +19d27'17''.8). 
SN~2014U is located 1''.0 west and 2''.8 north of the center of the galaxy 
NGC~3859 at $z_{\rm{NED}}=0.01824$ \citep{haynes1997}. 
Nothing is seen at this position in an image taken on Jan. 31.8, 2014. 
A near-infrared spectrum was taken with the FoldedPort Infrared Echellette (FIRE)
 spectrograph on the 6.5-m Magellan Baade Telescope on Feb. 27.2, 2014. 
The near-infrared spectrum reduced with basic methods \citep{hsiao2013} 
is similar to that of SN~2013hj at approximately 21 days past explosion 
with several hydrogen Paschen P-Cyg lines, indicating that SN~2014U is a Type~II SN. 

\item	SN~2014an\\
SN~2014an ($g=18.6$) was discovered on Mar. 31.7, 2014  
at a position (14h51m42s.93, +08d34'12''.5). 
SN~2014an is located 5''.3 east and 15''.8 north of the center of the galaxy 
SDSS~J145142.57+083356.6 (CGCG~076-079) at $z_{\rm{SDSS}}=0.060$. 
Nothing is seen at this position in an image taken on Feb. 24.7, 2014. 
This SN was confirmed to be $V=18.9$ with the Swope 1-m telescope at Las Campanas Observatory on Apr. 2.8, 2014. 
An optical spectrum of SN~2014an was taken with KOOLS on the Okayama188-cm telescope on Apr. 2.7, 2014. 
Cross-correlation with the SNID code shows that the best match is a normal SN~Ia, SN~1984A, at $t=-6$~days. 

\item	SN~2014bd\\
SN~2014bd ($g=20.0$) was discovered on Apr 23.6, 2014 
at a position (14h50m48s.07, +09d22'48''.3). 
SN~2014bd is located 1''.4 west and 5''.9 north of 
the center of the galaxy SDSS~J145048.16+092242.3 at $z_{\rm{SDSS}}=0.029$. 
Nothing is seen at this position in the image taken on Mar. 31.7, 2014. 
The object was confirmed with the Swope 1-m
telescope at Las Campanas Observatory on Apr 24.2, 2014 at $r=19.8$.
An optical spectrum of SN~2014bd was obtained on May 6.3, 2014 with 
the WFCCD on the Las Campanas 2.5-m du Pont telescope. 
No order-sorting filters were used but our spectroscopic classification 
would not suffer due to this \citep{folatelli2013}, which is also the case for SN~2014bd. 
An H$\alpha$ P-Cygni profile is seen indicating that SN~2014bd is a type-II SN. 
Cross-correlation with the SNID code shows good matches with Type~IIP SN around 10~days after maximum 
and the best match is SN~2005cs at $t=13$~days. 

\item	SN~2014bo\\
SN~2014bo ($g=19.3$) was discovered on on May 19.7, 2014 at (16h27m46s.15, Decl. = +41d44'23".7). 
SN~2014bo is located 3''.4 east and 5''.0 north of the center of the host galaxy SDSS~J162745.84+414418.6 at $z_{\rm{SDSS}}=0.133$. 
Nothing is seen at this position in an image taken on May 17.5, 2014 (limiting mag 20.0).
The SN was confirmed in $R$-band images taken with the 0.7-m telescope at the Sternberg Astronomical Institute on May 20.9, 2014.
A spectrum of SN~2014bo was taken on May 27.5, 2014 with KOOLS on the Okayama 188-cm telescope. 
Spectral fitting with the SNID code indicates that the best-matched template spectrum to the SN~2014bo spectrum 
is a normal SN~Ia, SN~1994S, at 3 days before maximum. 
We note that the expected maximum absolute magnitude in $g$-band without K-correction could be as bright as $-20.0$, 
possibly indicating the overluminous nature of this SN~Ia. 
\end{itemize}


\clearpage


\begin{table*}
  \caption{
  A list of supernovae discovered and identified by KISS and reported to CBET except for SN~2014bk. 
  References are 
  [1] \citet{morokuma2012a},  [2] \citet{itoh2012}, 
  [3] \citet{morokuma2012b},  [4] \citet{walker2012a}, 
  [5] \citet{morokuma2012c},  [6] \citet{walker2012b},  
  [7] \citet{tanaka2013a},    [8] \citet{stritzinger2013a}, 
  [9] \citet{tanaka2013b},    [10] \citet{taddia2013a}, 
  [11] \citet{tanaka2013c},   [12] \citet{taddia2013b}, 
  [13] \citet{morokuma2013},  [14] \citet{walker2013}, 
  [15] \citet{matsumoto2013}, [16] \citet{stritzinger2013b}, 
  [17] \citet{morokuma2014a}, [18] \citet{walker2014a},  
  [19] \citet{tanaka2014a},   [20] \citet{walker2014b},  
  [21] \citet{tanaka2014b},   [22] \citet{walker2014c},  
  [23] \citet{tanaka2014c},   [24] \citet{hsiao2014},  
  [25] \citet{tominaga2014a}, [26] \citet{tanaka2014d},  
  [27] \citet{tominaga2014b}, [28] \citet{morrell2014},  
  [29] \citet{tominaga2014c}, [30] \citet{tanaka2014e}. 
  }\label{tab:kisssnlist}
  {\footnotesize
  \begin{center}
    \begin{tabular}{lclrrrrrllc}
      \hline
      {CBET name}& {SN type} & {best-match} & {$z$} & {$t_{\rm{disc}}$} & {$t_{\rm{spec}}$} & {$m_{\rm{disc}}$} & {$M_{\rm{peak}}$}  & {disc. date} & {spec. date} & {reference}\\\hline
      \hline
      SN~2012cm & Ia  & SN~2002hu & 0.028 & $-8$ & $-4$ & 16.7 & $-19.2$ & 2012/05/13.6 & 2012/05/17.5 & [1],[2]\\
      SN~2012cq & IIP & SN~2004et & 0.026 & $-5$ & $+6$ & 17.4 & $-17.9$ & 2012/05/13.5 & 2012/05/24.9 & [3],[4]\\
      SN~2012ct & IIP & SN~2005cs & 0.039 & $-2$ & $0$  & 19.4 & $-16.8$ & 2012/05/22.5 & 2012/05/24.9 & [5],[6]\\
      SN~2013I  & IIP & SN~2005cs & 0.035 & $+1$ & $+5$ & 18.7 & $-17.1$ & 2013/01/11.4 & 2013/01/15.9 & [7],[8]\\
      SN~2013J  & Ic  & SN~1994I  & 0.034 & $-3$ & $-2$ & 18.7 & $-16.9$ & 2013/01/19.8 & 2013/01/21.2 & [9],[10]\\
      SN~2013Y  & Ia  & SN~2005cf & 0.077 & $0$  & $+5$ & 18.6 & $-18.9$ & 2013/02/06.6 & 2013/02/11.2 & [11],[12]\\
      SN~2013al & Ia  & SN~2006cf & 0.132 & $0$  & $+4$ & 19.3 & $-19.6$ & 2013/03/03.4 & 2013/03/07.1 & [13],[14]\\
      SN~2013ba & Ia  & SN~1997bp & 0.08\footnotemark[$*$] & $-5$ & $-3$ & 19.2 & $-18.7$ & 2013/04/06.7 & 2013/04/07.2 & [15],[16]\\
      SN~2014Q  & Ia  & SN~2002he & 0.046 & $-11$ & $-3$ & 18.9 & $-18.3$ & 2014/01/29.4 & 2014/02/06.0 & [17],[18]\\
      SN~2014S  & IIP & SN~1999gi & 0.04\footnotemark[$*$] & $-8$ & $-4$ & 18.6 & $-18.2$ & 2014/02/21.5 & 2014/02/25.1 & [19],[20]\\
      SN~2014T  & IIn & SN~1996L  & 0.09\footnotemark[$*$] & $+5$ & $+8$ & 18.7 & $-19.2$ & 2014/02/22.7 & 2014/02/25.2 & [21],[22]\\
      SN~2014U  & II  & SN~2013hj & 0.018 & $+17$ & $+21$ & 18.4 & $-16.4$ & 2014/02/23.5 & 2014/02/27.2 & [23],[24]\\
      SN~2014an & Ia  & SN~1984A  & 0.060 & $-8$  & $-6$  & 18.6 & $-18.7$ & 2014/03/31.7 & 2014/04/02.7 & [25],[26]\\
      SN~2014bd & IIP & SN~2005cs & 0.029 & $0$   & $+13$ & 20.2 & $-15.5$ & 2014/04/23.6 & 2014/05/06.3 & [27],[28]\\
      SN~2014bo & Ia  & SN~2004L  & 0.133 & $-11$ & $-3$  & 19.1 & $-20.0$ & 2014/05/19.7 & 2014/05/27.5 & [29],[30]\\
      \hline
        \multicolumn{4}{@{}l@{}}{\hbox to 0pt{\parbox{85mm}{\footnotesize
       \par\noindent
       \footnotemark[$*$] These redshifts are obtained by fitting the SN spectra. 
     }\hss}}
    \end{tabular}
  \end{center}
  }
\end{table*}

\begin{table*}
  \caption{
  A list of spectroscopic observations for the 
  SNe reported to CBET except for SN~2014bk. 
  }\label{tab:specobs}
  {\footnotesize
  \begin{center}
    \begin{tabular}{llclccll}
      \hline
      {telescope} & {instrument} & {slit }  & {grism} & {$R$} & {$\lambda$ range}  & {$t_{\rm{exp}}$}\\
      {} & {} & {width}  & {} & {} & {[A]}  & {[sec]}\\\hline
      \hline
      Kanata & HOWPol    & 2.2 & 420~l/mm & 400 & 4500-9000 & $300\times6$ (SN~2012cm)\\
      TNG     & DOLORES & 1.5 & LR-B        & 400 & 3500-8000 & $1200\times2$ (SN~2012cq, SN~2012ct, SN~2013al), $900\times2$ (SN~2014Q)\\
      & & 1.0 & LR-B  & 600 & 3500-8000 & $900\times2$ (SN~2014S), $1200\times2$ (SN~2014T)\\
      NOT     & ALFOSC    & 1.0 & \#4~300~l/mm & 350 & 3500-8000 & $2400\times1$ (SN~2013I, SN~2013J, SN~2013Y, SN~2013ba)\\ 
      Magellan & FIRE       & 0.6     & prism               & 500  & 8100-24000 & $126.8\times8$ (SN~2014U)\\
      OAO188  & KOOLS  & 1.8     & \#5                   & 500  & 4000-7400   & $1800\times1$ (SN~2014an), $1800\times4$ (SN~2014bo)\\
      du Pont    & WFCCD & 1.65  & blue 400~l/mm & 750 & 3800-9200    & $1500\times2$ (SN~2014bd)\\
      \hline
        \multicolumn{4}{@{}l@{}}{\hbox to 0pt{\parbox{85mm}{\footnotesize
     }\hss}}
    \end{tabular}
  \end{center}
  }
\end{table*}



\end{document}